\documentclass[aps,twocolumn]{revtex4}
\usepackage{psfig} 

\newcommand{\gtsimeq}{\raisebox{-0.6ex}{$\,\stackrel
        {\raisebox{-.2ex}{$\textstyle >$}}{\sim}\,$}}

\begin{document} 



\title{Separating internal and external  dynamics of complex systems}
\author{M. Argollo de Menezes and A.-L. Barab\'asi}
\affiliation{Department of Physics, University of Notre Dame, Notre Dame,
IN 46556}
\date{\today}
\begin{abstract}
  { The observable behavior of a complex system reflects the
    mechanisms governing the internal interactions between the
    system's components and the effect of external perturbations.
    Here we show that by capturing the simultaneous activity of
    several of the system's components we can separate the internal
    dynamics from the external fluctuations. The method allows us to
    systematically determine the origin of fluctuations in various
    real systems, finding that while the Internet and the computer
    chip have robust internal dynamics, highway and Web traffic are
    driven by external demand. As multichannel measurements are
    becoming the norm in most fields, the method could help uncover
    the collective dynamics of a wide array of complex systems.  }
\end{abstract}
\maketitle


Decades of research has lead to the development of sophisticated tools
to analyze time series generated by various dynamical systems,
allowing us to extract short and long range temporal correlations,
periodic patterns and stationarity information
\cite{abarbanel,dfa,havlin2003,non-stationary}.  We lack, however,
systematic methods to extract from multiple datasets information not
already provided by a single time series. Indeed, advances in computer
aided measurement techniques increasingly offer the possibility to
separately but simultaneously record the time dependent activity of a
system's many components, such as information flow on thousands of
Internet routers or highway traffic on numerous highways.  As the time
dependent activity of each component (router or highway) captures the
system's dynamics from a different angle, these parallel time series
offer us increasingly complete information about the system's
collective behavior. Yet, we have difficulty answering a simple
question: How can we uncover from multiple time series a system's
internal dynamics?

Multiple time series are typically available for complex systems whose
dynamics is determined by the interaction of a large number of
components that communicate with each other through some complex
network ~\cite{alb-rmp}. The dynamics of each component is
determined by two factors: (1) interactions between the components,
governed by some internal dynamical rules that distribute the activity
between the various parts of the system and (2) global variations in
the overall activity of the system.  For example, the traffic increase
on highways during peak hours and surges in the number of Internet
users during working hours represent global activity changes that have
a strong impact on the local activity of each component (highway or
router) as well.  Different components are influenced to a different
degree by these global changes, making impossible for an observer that
has access only to a single component to separate the internal
dynamics from the externally imposed fluctuations.  Most important,
the inevitable fluctuations in the external conditions systematically
obscure the mechanisms that govern the system's internal dynamics.

Here we propose a method to separate in a systematic manner for each
time series the external from the internal contributions, and validate
it on model systems, for which the magnitude of the external
perturbations can be explicitly controlled.  By removing the impact of
the external changes on the system's activity we gain insights into
the internal dynamics of a wide range of systems, from Internet
traffic to bit flow on a microprocessor.

Let us consider a dynamical system for which we can record the time
dependent activity of $N$ components, allowing us to assign to each
component $i$ a time series $\{f_i(t)\}$, $t=1,\ldots,T$ and
$i=1,\ldots,N$. As each time series reflects the joint contribution
from the system's internal dynamics and external fluctuations, we
assume that we can separate the two contributions by writing

\begin{equation}
f_i(t)=f_i^{int}(t)+f_i^{ext}(t).
\label{eq:separation}
\end{equation}

To determine $f_i^{ext}(t)$ let us consider the case when internal
fluctuations are absent, and therefore the total traffic in the system
is distributed in a deterministic fashion among all components.  In
this case component $i$ captures a time independent fraction $A_i$ of
the total traffic.  For different components $i$, $A_i$ can differ
significantly, being determined by the component's centrality
\cite{heiko}.  The challenge is to extract $A_i$ from the
experimentally available data without knowledge of the system's
internal topology or the dynamical rules governing its activity.  For
this we write $A_i$ as the ratio of the total traffic going through
the component $i$ in the time interval $t \in [0,T]$ and the total
traffic going over all observed components during the same time
interval

\begin{equation}
A_i = \frac{\sum_{t=1}^{T} f_i(t)}{\sum_{t=1}^T\sum_{i=1}^{N} f_i(t)}.
\label{eq:a_i}
\end{equation}

\noindent  At any moment $t$ the amount of traffic {\it expected} to go
through node $i$ is therefore given by the product of $A_i$ and the
total traffic in the system in moment $t$ (i.e. $\sum_{i=1}^Nf_i(t)$),
providing the magnitude of the traffic expected if only external
fluctuations contribute to the activity of node $i$ as
\begin{equation}
f^{ext}_i(t)=A_i\sum_{i=1}^Nf_i(t).
\label{eq:external}
\end{equation}

\noindent Equation (\ref{eq:external}) describes the case in
which changes in the system's overall activity are reflected in a
proportional fashion on each component. Real systems do display,
however, internal fluctuations, which will generate local and temporal
deviations from the expected $f_i^{ext}(t)$, a consequence of the
internal time dependent redistribution of traffic in the system. Using
$(1-3)$ we obtain this internal component as

\begin{equation}
f_i^{int}(t) = f_i(t) - \left(\frac{\sum_{t=1}^{T}
f_i(t)}{\sum_{t=1}^T\sum_{i=1}^{N} f_i(t)}\right) \sum_{i=1}^Nf_i(t),
\label{eq:internal}
\end{equation}

\noindent which, by definition, has zero average, as it captures 
the deviations from the traffic expected to go through component $i$.
Given the experimentally measured dynamic signal $f_i(t)$ on a large
number of components, (\ref{eq:external}) and (\ref{eq:internal})
allow us to separate each signal $f_i(t)$ into two contributions
$f_i^{ext}(t)$ and $f_i^{int}(t)$, the first capturing changes in the
system's overall activity, providing a measure of the external
fluctuations and the second describing the fluctuations characterizing
the system's internal dynamics.

\begin{figure}[!ht]
  {\centerline{\psfig{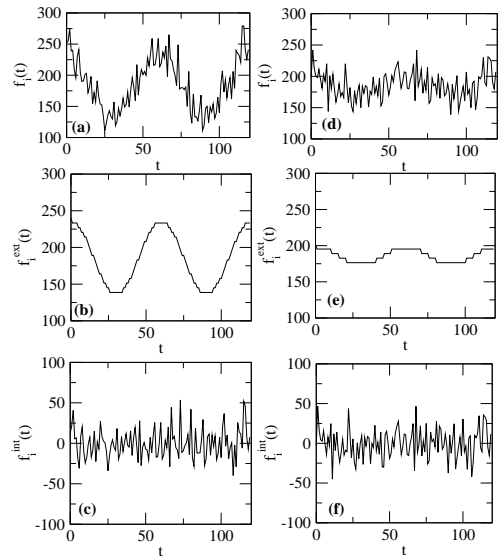}}}
  \caption{Splitting a measured signal into its external and internal
    contributions for the model system. We study the traffic on the
    nodes of a scale-free network with $10^3$ nodes, generated by
    $M(t)$ random walkers whose number follow a sinusoidal signal with
    amplitude $\Delta M$ and period $60$.  {\bf (a)} The activity
    measured on a typical node when the amplitude of the external
    fluctuations is $\Delta M=1000$.  {\bf (b)} The external
    contribution provided by Eq. (\ref{eq:external}) recovers the
    periodic signal imposed on the system. {\bf (c)} The internal
    contribution, predicted by Eq. (\ref{eq:internal}), captures the
    random pattern of the diffusion process. {\bf (d-f)} The same as
    in {\bf (a-c)}, but with a small external amplitude $\Delta M=10$,
    demonstrating that the method works even when the amplitude of
    external fluctuations are comparable to fluctuations of the
    internal dynamics.}
\label{signals-model}
\end{figure}

To test the ability of (\ref{eq:external}) and (\ref{eq:internal}) to
separate the internal and external components of a time series we
investigate a simple model system of random walkers on a network
\cite{us}.  We randomly displace $M(t)$ non-interacting walkers on the
network, allowing each to perform $N_s$ steps and monitor the total
number of visitations $f_i$ for each node $i$.  If we repeat the
experiment $T$ times, we find that the number of visits to node $i$
differs from one experiment to the other, the time series
$\{f_i(t)\},~ t=1\ldots T$ characterizing the fluctuations intrinsic
to the diffusion process.  If, however, we allow the number of walkers
$M(t)$ to vary from one experiment to the other, the local variations
in $f_i(t)$ are rooted not only in the random character of diffusion,
but also in variations imposed by changes in the total activity
$M(t)$.  An observer that records only a single ${f_i(t)}$ time series
has difficulty deciding if the measured fluctuations reflect the
system's internal dynamics only, or some non-stationary external
effect.  To test the method's ability to separate the internal and
external fluctuations we use an external signal with an easily
recognizable periodic profile $M(t)=\left<M\right> + \Delta
M\sin(kt)$.  Figs. \ref{signals-model}a and \ref{signals-model}d show
the activity $f_i(t)$ recorded for a typical node for two different
$\Delta M$ amplitudes, representing a visible superposition of the
sinusoidal external signal and the internal randomness of the
diffusion process.  As Figs.\ref{signals-model}b and
\ref{signals-model}e show, the external component provided by
(\ref{eq:external}) fully recovers the external signal imposed on the
system. After removing the external component using
Eq.(\ref{eq:internal}) we obtain a random pattern reflecting the
intrinsic fluctuations of the diffusion process.  The method works
equally well in the case when the magnitude of the external
fluctuations is large (Figs.\ref{signals-model}a) or small
(Figs.\ref{signals-model}d) compared to the system's internal
fluctuations \cite{incomplete}.

\noindent We can use (\ref{eq:external}) and  (\ref{eq:internal}) to 
determine if the fluctuations observed in a system are mainly internal
or externally imposed.  For each recorded signal $i$ we determine the
external and internal standard deviations, $\sigma_i^{ext} =
\sqrt{\left<f_i^{ext}(t)^2\right> - \left<f_i^{ext}(t)\right>^2}$ and
$\sigma_i^{int} = \sqrt{\left<f_i^{int}(t)^2\right> -
  \left<f_i^{int}(t)\right>^2}$, and their ratio
\begin{equation}
\eta_i = \frac{\sigma_i^{ext}}{\sigma_i^{int}}.
\end{equation}

\begin{figure}[!htbp]
  {\centerline{\psfig{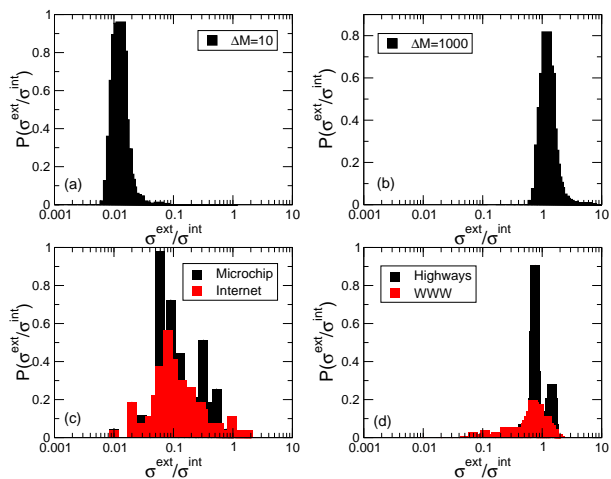}}}
  \caption{{\bf}  Distribution of $\eta_i = \sigma_i^{ext}/\sigma_i^{int}$ ratios
    of external and internal fluctuations for model {\bf (a,b)} and
    selected real systems {\bf (c,d)}.  Distribution of the $\eta_i$
    ratios for the random walk model: {\bf (a)} For smaller external
    fluctuations, the distribution is centered around a small value of
    $\eta \ll 1$, indicating that internal fluctuations overcome
    external ones, dominating the system's dynamics. {\bf (b)} When
    $\Delta M$ is increased, however, such fluctuations overshadow the
    system's internal dynamics, and the $P(\eta)$ distribution shifts
    towards larger values of $\eta$ (right curve).  {\bf (c)}
    $P(\eta)$ distributions for the Internet and the microchip,
    centered around $\eta \sim 0.1$, indicate that external
    fluctuations do not affect the system's overall dynamics
    significantly. {\bf (d)} The World Wide Web and the Highway
    networks, with $P(\eta)$ peaked around $\eta\sim 1$, are strongly
    influenced by fluctuations in the total number of web surfers and
    the number of cars, respectively. }
\label{fig:eta-all}
\end{figure}

\noindent When $\eta_i \gg 1$ the external
fluctuations dominate the dynamics of component $i$, while for $\eta_i
\ll 1$ the system's internal dynamics dominates over the externally
imposed changes.  As different signals have different $\eta_i$ values,
the system's overall behavior is best characterized by the $P(\eta_i)$
distribution, obtained after calculating $\eta_i$ for each signal we
have access to. Figures \ref{fig:eta-all}a and b show $P(\eta_i)$ for
the random walk model, in which the number of walkers follows
$M(t)=\left<M\right> +\xi_i(t)$ where $\xi_i(t)$ is a random variable
uniformly distributed between $-\Delta M/2$ and $\Delta M/2$.  For
small external fluctuations ($\Delta M \simeq 0$) the $P(\eta_i)$
distribution is highly peaked and is located entirely in the $\eta_i
\ll 1$ region, indicating that external fluctuations have little
influence on the dynamics of the individual components.  For high
$\Delta M$, however, $P(\eta_i)$ lies in the $\eta \gtsimeq 1$ region,
indicating that the system's dynamics is dominated by external
fluctuations.

Our ability to split the time series into an internal and external
signal offers novel insights into the dynamics of four systems of
major technological importance: Internet routers \cite{vazquez}, a
microchip \cite{sole}, the World-Wide-Web \cite{lawrence} and the
highway system of Colorado. We collected time resolved information
about the activity of a large number of components, such as traffic on
$374$ Internet routers, switching behavior of $462$ gates of a
microchip, the daily visitations of $3000$ web sites on the Web and
the daily traffic for $127$ highways in Colorado (details about the
databases are provided in Ref.  \cite{us}). We used $(1-4)$ to
separate the signal for each component $i$, the corresponding
$P(\eta)$ distribution unraveling clear differences between the
studied systems.  We find that for the Internet and the microchip
internal fluctuations dominate over the externally induced changes, as
the $P(\eta)$ distribution lies in the $\eta \ll 1$ region, peaked
around $\eta \simeq 0.08$ (Fig.\ref{fig:eta-all}c).  On the other
hand, for the World Wide Web and highways the typical $\eta$ ratios
are an order of magnitude larger (Fig.  \ref{fig:eta-all}d), the
$P(\eta)$ distribution being peaked at $\eta \simeq 1$, indicating
that for these two systems the external and internal fluctuations are
comparable in magnitude.

\noindent This  separation correlates with the  finding that
the four studied systems belong to two distinct universality classes
\cite{us}. Indeed, for each recorded signal the time average
$\left<f_i\right>$ and the standard deviation $\sigma_i$ obey the
scaling law $\sigma_i \sim \left<f_i\right>^{\alpha}$, where for the
microchip and the Internet $\alpha=1/2$, while for the highways and
the WWW $\alpha=1$.  Figure \ref{fig:eta-all} indicates that $\alpha$
correlates with the relative magnitude of the external fluctuations
\cite{us}: for systems with $\alpha=1/2$ the internal fluctuations
dominate (Fig.  \ref{fig:eta-all}c), while for systems with $\alpha=1$
the impact of the external fluctuations are at least comparable to the
fluctuations generated by the system's internal dynamics (Fig.
\ref{fig:eta-all}d).

\begin{figure}[!ht]
\vskip 0.5cm
  {\centerline{\psfig{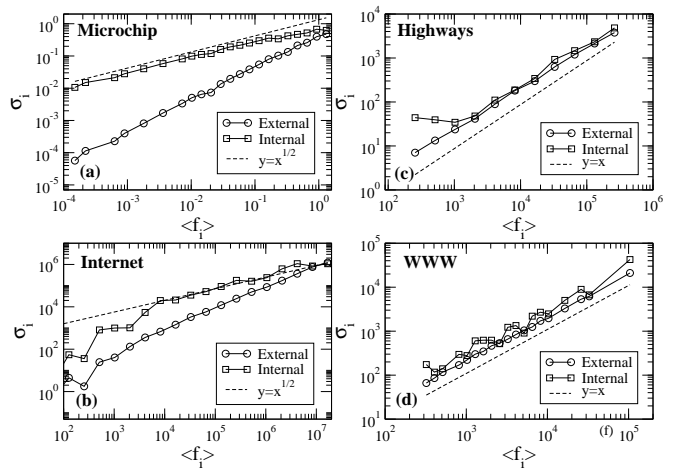}}}
  \caption{Scaling of the external and internal fluctuations with the
    average flux. Internal fluctuations $\sigma_i^{int}$ on the
    microchip {\bf (a)} and on the Internet {\bf (b)}, both belonging
    to the $\alpha=1/2$ class, are significantly larger than external
    fluctuations $\sigma_i^{ext}$, and scale with a different
    exponent.  External and internal fluctuations are comparable in
    magnitude on the World Wide Web {\bf (c)} and the highway network
    {\bf (d)}, and they also follow the same scaling, indicating that
    in these systems external fluctuations should have strong impact
    on systems' overall dynamics.}
\label{fig:scaling-exp}
\end{figure}

The $P(\eta)$ distribution tells us only the origins of the
fluctuations, and is not sufficient to understand the intimate
differences between the internal and the external contributions.  A
more detailed understanding is provided by plotting for each signal
$i$ the $\sigma_i^{ext}$ and the $\sigma_i^{int}$ standard deviations
in function of the average $\left<f_i\right>$ (Fig.
\ref{fig:scaling-exp}a-d). We find that for the microchip and the
Internet $\sigma_i^{ext}$ and $\sigma_i^{int}$ scale with different
exponents (Fig.  \ref{fig:scaling-exp}a,b): the internal fluctuations
scale with $\alpha=1/2$, while the external signal scales with
$\alpha=1$ (which is an expected feature of the external fluctuations
\cite{us}).  Furthermore, for these two systems the internal standard
deviation is much larger than the external one
($\sigma_i^{int}(\left<f_i\right>) \gg
\sigma_i^{ext}(\left<f_i\right>)$, explaining why the overall $\sigma
\sim \left<f\right>^{\alpha}$ scaling captures only the $\alpha=1/2$
exponent.  In contrast, for the WWW and the highways the
$\sigma_i^{ext}$ and $\sigma_i^{int}$ curves overlap, both following
the $\alpha=1$ exponent (Fig.  \ref{fig:scaling-exp}c,d).

\noindent The qualitative difference between the two sets of plots in
Fig. \ref{fig:scaling-exp} reflect fundamental differences in the
internal dynamics of the four studied real systems. The splitting of
the curves seen in Figs. \ref{fig:scaling-exp}a and b indicates that
the Internet and microchip are characterized by a robust internal
dynamics, that leads to a dominating $\alpha=1/2$ internal scaling.
While the $\alpha=1/2$ exponent emerges in the studied diffusion model
as well \cite{us}, the nature of the internal dynamics and the origin
of the $1/2$ exponent needs to be addressed in each system separately.
In contrast, the overlapping curves seen in Fig.
\ref{fig:scaling-exp}c,d indicate not only that highway and WWW
traffic are much more susceptible to external perturbations, but
suggest that these systems do not have a clearly separable internal
dynamics. That is, the local activity of the system is driven simply
by global demand, and the interactions between the various highways or
web sites do not lead to a distinguishable internal dynamics.  Indeed,
while on the microchip and the Internet there are strict protocols
regulating the traffic of bits or packages, highways and the WWW allow
for a much higher flexibility, the users having the option to leave
the system each time they encounter unforeseen local conditions, like
highway congestion or Web delays. Yet, highway and Internet traffic in
many ways are quite similar \cite{debabish,csabai}, each describing a
clear source-destination shortest-path traffic. Thus, the fundamental
difference in their internal dynamics is in many ways surprising and
warrants further inquiry.

Our simulations indicate that a non-stationary external noise does not
affect the method's applicability, as the non-stationary behavior will
be carried by the external component of the separated signal.
However, it is unclear if the method could be applied if there is
internal nonstationarity in the system, corresponding to time
dependent shifts in the system's overall activity between groups of
nodes. Such internal nonstationarity can be uncovered by calculating
the $A_i$ parameters in non-overlapping time windows \cite{tass},
potentially resulting in significant lasting shifts in the $A_i$
values. An inspection of the four studied systems did not reveal
non-stationary internal behavior, the $A_i$ parameter fluctuating
around $\left<A_i\right>$. The method appears to be insensitive to the
choice of the observational window $T$ used in Eq.  $(2)$, as long as
$T$ is large enough so that the average can be evaluated.

In an increasing number of complex systems one can experimentally
monitor the simultaneous activity of hundreds of channels, examples
including multichannel measurement of neural activity on {\it in vivo}
cell colonies \cite{neural_chip}, simultaneous monitoring of thousands
of gene expression data sets for model organisms, like {\it E.  Coli}
or {\it S. Cerevisae} \cite{holter}, flow fluctuation in river
networks \cite{banavar}, price variations in individual stocks or
goods \cite{stanley} or the activity of different processors in
parallel computation \cite{korniss}.  The method introduced here
represents a systematic tool for extracting information from multiple
channel measurements, offering detailed insights into the mechanisms
that govern the dynamics of these systems.

Acknowledgments: We are indebted to Jay Brockman and Paul Balensiefer
for providing data on the computer chip, and to J\'anos Kert\'esz for
fruitful discussions.  This research was supported by grants from NSF,
NIH and DOE.


\begin{thebibliography}{99}

  
\bibitem{abarbanel} H. Kautz and T.  Schreiber, {\it Nonlinear Time
    Series Analysis} (Cambridge Univ.  Press, Cambridge, 1997); H.D.I.
  Abarbanel, R. Brown, J. Sidorowich and L.  Tsimring, Rev. Mod. Phys.
  {\bf 65}, 1331 (1993).
  
\bibitem{dfa} C.-K. Peng , S.V. Buldyrev, S. Havlin, M. Simons, H.E.
  Stanley, A.L. Goldberger, Phys. Rev. E {\bf 49}, 1685-1689 (1994).
  
\bibitem{havlin2003} V. N. Livina, Y. Ashkenazy, P. Braun, R. Monetti,
  A. Bunde, and S. Havlin, Phys. Rev. E 67 , 042101 (2003).
    
\bibitem{non-stationary} C.-K. Peng, S. Havlin, H.E. Stanley, and Ary
  L. Goldberger, Chaos {\bf 5}, 82 (1995); J.W.  Kantelhardt, S.A.
  Zschiegner, E. Koscielny-Bunde, A. Bunde, S.  Havlin, H.E. Stanley,
  Physica A {\bf 316}, 87 (2002).

  
\bibitem{alb-rmp} R. Albert and A.-L. Barab\'asi, Rev. Mod. Phys. {\bf
    74}, 47 (2002); S.N. Dorogovtsev, J.F.F. Mendes, {\it Evolution of
    Networks: From Biological Nets to the Internet and WWW} (Oxford
  University Press, Oxford, 2003).
  
\bibitem{heiko} J.D. Noh and H. Rieger, Phys. Rev. Lett. {\bf 92},
  118701 (2004).
  
\bibitem{incomplete} For most real systems we can monitor only a small
  fraction of the components.  To show that the method works for
  incomplete datasets as well, we applied $(1-4)$ to signals recorded
  only from a randomly chosen fraction of nodes, again successfully
  recovering the external signal.
  
  
\bibitem{vazquez} A. Vazquez, R. Pastor-Satorras, A. Vespignani, {\it
    Phys. Rev. E} {\bf 65}, 066130 (2002); S.-H. Yook, H. Jeong, and
  A.-L. Barab\'asi, Proc. Natl. Acad. Sci. {\bf 99}, 13382 (2002).
  
\bibitem{sole} R. F. Cancho, C. Janssen, R.V. Sole, {\it Phys. Rev E},
  {\bf64}, 046119 (2001).

\bibitem{lawrence} S. Lawrence, L. Giles, {\it Science} {\bf 280},
  98-100 (1998); R. Albert, H. Jeong, and A.-L. Barab\'asi, Nature
  {\bf 401}, 130 (1999).
  
\bibitem{us} M. Argollo de Menezes and A.-L. Barab\'asi, Phys. Rev.
  Lett. {\bf 92}, 028701 (2004).
  
\bibitem{kilpatrick} A.M Kilpatrick and A.R. Ives, Nature {\bf 422},
  65-68 (2003).

\bibitem{debabish} D. Chowdhury, L. Santen, and A. Schadschneider,
  Physics Reports {\bf 329}, 199 (2000).
  
\bibitem{csabai} G. Simon and I. Csabai, Physica {\bf A 307}, 516-526
  (2002).
  
\bibitem{tass} P. Tass, M.G. Rosenblum, J. Weule, J. Kurths, A.
  Pikovsky, J. Volkmann, A. Scnhitzler and H.-J. Freund, Phys. Rev.
  Lett. {\bf 81}, 3291 (1998); F. Rieke, D.  Warland, R.R. van Steveninck and W.
  Bialek, {\it Spikes: Exploring the Neural Code}, (MIT Press,
  Massachusetts, 1999); C.-K. Peng, S.V. Buldyrev, S. Havlin, M. Simons, H.E.
  Stanley and A.L Goldberger, Phys. Rev. E {\bf 49}, 1685 (1994).
  
\bibitem{neural_chip} M.P. Maher, J. Pine, J. Wright and Y.-C. Tai, J.
  Neurosc. Methods {\bf 87}, 45-56 (1999).

\bibitem{holter} N.S. Holter, A. Maritan, M. Cieplak, N.Fedoroff,
J.R. Banavar, {\it Proc. Natl. Acad. Sci. USA} {\bf 98} 1693-1698
(2001); J. Hasty, J.J. Collins, {\it Nat.  Genet.} {\bf 31}
13-14 (2002).

\bibitem{banavar} V. N. Livina, Y. Ashkenazy, P. Braun, R. Monetti, A.
  Bunde, S. Havlin Phys. Rev. E {\bf 67}, 042101 (2003); J.R. Banavar,
  A. Maritan, A. Rinaldo, {\it Nature} {\bf 399}, 130-132 (1999); M.
  Cieplak, A. Giacometti, A. Maritan, A.  Rinaldo, J.R. Banavar, {\it
    J. Stat. Phys.}  {\bf 91}, 1-15 (1998); G. Caldarelli, {\it Phys.
    Rev. E} {\bf 63}, 21118 (2001).

\bibitem{stanley} R.N. Mantegna, H.E. Stanley, {\it An Introduction to
  Econophysics: Correlations and Complexity in Finance} (Cambridge
  Univ. Press, New York, 2000).

\bibitem{korniss} G. Korniss, N.A. Novotny, H. Guclu, Z. Toroczkai,
P.A Rikvold, {\it Science} {\bf 299}, 677-679 (2003).
 \end{thebibliography}
\end{document}